\documentclass[aps,reprint,superscriptaddress,longbibliography]{revtex4-1}
\usepackage{blindtext}

\usepackage{amsmath, amssymb, mathtools, bm, nicefrac, physics}

\usepackage{csquotes}

\usepackage{graphicx}
\usepackage[caption=false]{subfig}

\usepackage[colorlinks=true, allcolors=monbleu]{hyperref}

\usepackage{xcolor}
\definecolor{myTeal}{RGB}{0,135,155}	

\definecolor{myGreen}{RGB}{67,111,38}	

\definecolor{myOrange}{RGB}{168,84,0}	

\definecolor{monbleu}{RGB}{76, 114, 176}

\def\Vol{\mathrm{Vol}}

\graphicspath{{figures/}{figures_ps/}}

\begin{document}
\title{Dimension matters when modeling network communities in hyperbolic spaces}
\date{\today}
\author{B\'eatrice D\'esy}
\email{beatrice.desy@vuw.ac.nz}
\affiliation{School of Information Management, Victoria University of Wellington, Wellington, New Zealand}
\affiliation{Antarctic Research Centre, Victoria University of Wellington, Wellington, New Zealand}

\author{Patrick Desrosiers}
\affiliation{D\'epartement de physique, de g\'enie physique et d'optique, Universit\'e Laval, Qu\'ebec, QC, Canada G1V 0A6}
\affiliation{Centre interdisciplinaire en mod\'elisation math\'ematique, Universit\'e Laval, Qu\'ebec, QC, Canada G1V 0A6}
\affiliation{Centre de recherche CERVO, Qu\'ebec, QC Canada G1J 2G3}

\author{Antoine Allard}
\affiliation{D\'epartement de physique, de g\'enie physique et d'optique, Universit\'e Laval, Qu\'ebec, QC, Canada G1V 0A6}
\affiliation{Centre interdisciplinaire en mod\'elisation math\'ematique, Universit\'e Laval, Qu\'ebec, QC, Canada G1V 0A6}

\begin{abstract}
Over the last decade, random hyperbolic graphs have proved successful in providing geometric explanations for many key properties of real-world networks, including strong clustering, high navigability, and heterogeneous degree distributions. These properties are ubiquitous in systems as varied as the internet, transportation, brain or epidemic networks, which are thus unified under the hyperbolic network interpretation on a surface of constant negative curvature. Although a few studies have shown that hyperbolic models can generate community structures, another salient feature observed in real networks, we argue that the current models are overlooking the choice of the latent space dimensionality that is required to adequately represent clustered networked data. We show that there is an important qualitative difference between the lowest-dimensional model and its higher-dimensional counterparts with respect to how similarity between nodes restricts connection probabilities. Since more dimensions also increase the number of nearest neighbors for angular clusters representing communities, considering only one more dimension allows us to generate more realistic and diverse community structures. 
\end{abstract}

\maketitle

\section{Introduction}
When one pictures a system and relationships between its constituents, the idea of closeness naturally comes to mind because more often than not, the flows that make up those relationships depend upon some form of proximity. Take neurons in a brain or servers underlying the infrastructure of the Internet, two systems that may seem to have little in common. Yet, both are remarkably complex and adequately modeled in the field of network geometry~\cite{Boguna2010, Allard2020, Boguna2021}, where closeness between nodes and properties of an underlying abstract space explain how elements of the network are interconnected. In particular, a latent space of constant negative curvature, the two-dimensional hyperbolic plane, captures in a simple yet accurate way many significant complex network properties, namely sparsity, self-similarity, small-worldness, heterogeneity, non-vanishing clustering, and community structure~\cite{Boguna2020}. 

In this framework, each node exists on a surface where a radial dimension encodes its popularity, or how likely it is to have many neighbors, and an angular dimension encodes the similarity between nodes as attractiveness, such that similar nodes are more likely to be related independently of their popularity~\cite{Papadopoulos2012}. The successes of hyperbolic network geometry cover a wide range of practical applications, like predicting economic patterns across time~\cite{Garcia-Perez2016}, making sense of the resilience of the Internet~\cite{Boguna2010} or modeling information flow in the brain~\cite{Allard2020, Zheng2021}, to name a few. Furthermore, hyperbolic space is the only known metric space on which maximum-entropy random graphs can reproduce real network properties like clustering, sparsity, and heterogeneous degree distributions all at once~\cite{Boguna2020}. The model has also been extended to weighted~\cite{Allard2017}, growing~\cite{Papadopoulos2012, Krioukov2012}, bipartite~\cite{Kitsak2017}, multilayer~\cite{Kleineberg2016, Kleineberg2017} or modular networks~\cite{Garcia-Perez2018, Zuev2015, Muscoloni2018}. 

Now that hyperbolic networks of the lowest dimension have been shown to capture so many realistic properties, some attention has shifted to the study of higher-dimensional models~\cite{Yang2020, Almagro2022, Kovacs2022, Kitsak2023}. In these, there is still one radial coordinate for popularity, but there are \(D>1\) dimensions encoding similarity, or perhaps, \emph{similarities}. In other words, higher-dimensional hyperbolic network models embody the intuition that there is more than one way in which things can be similar or not. The choice of dimension is an already prominent problem for machine learning applications of hyperbolic embeddings~\cite{Gu2021}, yet is has mostly been overlooked until recently for hyperbolic network models. In recent works, increasing the dimension was convoluted with the effect of other parameters and studied only at the local scale of node pairs connectivity and expected degrees~\cite{Yang2020, Kovacs2022, Kitsak2023}. These studies also found that similar power-law degree distributions can be achieved in any dimensions, by tuning the choice of model's parameter regime.

These observations involve extremely local properties, concerning nodes and their direct neighbors. As soon as we start zooming out towards the mesoscale level, dimension seems to impact network topology, yet it has not been studied much so far. The maximum clustering coefficient that can be achieved in random geometric graphs, which quantifies the closure of triplets of nodes into triangles, decreases with the dimension~\cite{Garcia-Perez2018b, Kovacs2022, Kitsak2023}. Almagro \emph{et al.}~\cite{Almagro2022} have shown using the short cycle structure that various networked data sets have an underlying hyperbolic dimension that is ultra-low, albeit not minimal as previously assumed. Our work follows this line of research by looking at this question through the lens of community structure.

One of the most ubiquitous properties of complex networks is community structure, when connectivity within subgroups of nodes, or communities, is prominently different than with the rest of the network~\cite{Girvan2002, Cherifi2019}. Network motifs have long been recognized as universal building blocks of complex networks~\cite{Milo2002}, community structure detection is one of the most long-standing active fields of network science~\cite{Fortunato2016,fortunato202220}, and most networked data show some sort of community structure, with one of the most interesting complex systems, the brain, making no exception~\cite{Betzel2017, Betzel2019}. In hyperbolic network models, community structure is expressed as subgroups of nodes that are closer with respect to their angular coordinates, either in hyperbolic embeddings of real networks~\cite{Boguna2010, Garcia-Perez2019} or synthetic models that explicitly generate communities~\cite{Zuev2015, Garcia-Perez2018, Muscoloni2018}. Since changing the dimension of hyperbolic network models affects primarily the number of angular coordinates, dimension and community structure are much closer than they appear. In our work, we explore this interplay to see how dimension can improve current modeling of community structure in the hyperbolic framework, but also how community structure offers some insight about the underlying hyperbolic dimension of networks.

\begin{figure}[h!]
\centering
\includegraphics[width=\columnwidth]{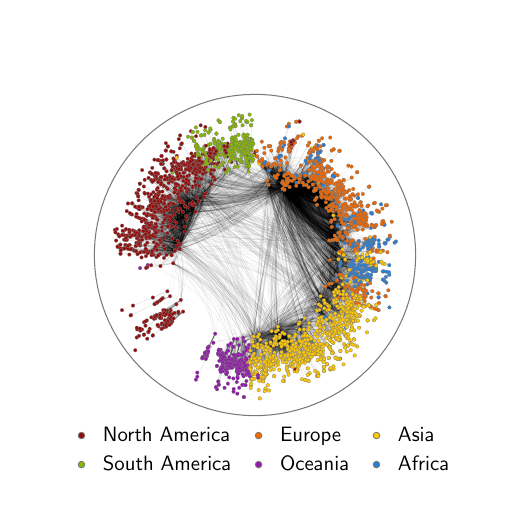}
\caption[Airports network embedded in the hyperbolic disk using the Mercator algorithm~\cite{Garcia-Perez2019}.]{Airports network embedded in the hyperbolic disk using the Mercator algorithm~\cite{Garcia-Perez2019}. Edges represent pairs of airports connected by a direct flight and colors represent the continent on which the airport is located. Here, edges follows hyperbolic geodesics of the conformal disk model~\cite{Ratcliffe2006}. Data downloaded from \url{openflights.org}. \label{fig:airports}}
\end{figure}

Let us take a look at a hyperbolic embedding of the airport network in Fig.~\ref{fig:airports}. Airports within the same continent are more prone to be connected by direct flights, which is why nodes of the same color are mostly grouped by angular coordinate. Nevertheless, airports in Africa, Asia, and Europe seem to be more mixed together because their actual geographic location has, very broadly, the shape of a triangle which cannot be reflected in the maximum likelihood embedding of Fig.~\ref{fig:airports}. This phenomenon is also notable in structural brain networks hyperbolic embeddings of Ref.~\cite{Allard2020}, where neuroanatomical clusters are not all grouped by angular coordinates. These examples illustrate that a unique similarity dimension might not be enough to model community structure with non-sequential patterns. If that were the case, one could wonder why is that so, and how can we quantify community structure in hyperbolic random graphs? 

As to why, we find that going from one angular similarity dimension to more, and even to only one more, have drastic effects on how the similarity between nodes influences their connectivity. We show that angular closeness constrains connections much more, and differently, in \(D=1\) than in other dimensions, which is done by studying the probability density function of the angular distance between connected nodes. We also quantify how the number of neighbors is not the same on a circle or on a sphere or on an even higher-dimensional sphere. This phenomenon impacts the number of nearest neighbors for angular clusters representing communities in hyperbolic random graphs. As a simplest experiment of how those two phenomenons come into play, we generate hyperbolic networks possessing community structure in \({D=1}\) and \({D=2}\). We thus obtain insights into how and why networks with community structure might have an underlying hyperbolic dimension that is higher than one.

The paper is structured as follows. In Sec.~\ref{sec:intro}, the key properties of the \(\mathbb{H}^{D+1}\) hyperbolic random graph model are recalled, along with its relationship to the \(\mathbb{S}^D\) formulation and some remarks about angular distance on \(D\)-spheres. The interplay between distance and dimension is studied in Sec.~\ref{sec:dimension}, where we will see how dimension affects how connected nodes in hyperbolic random graphs are likely to be found at a given angular distance from one another. In either the pairwise case, where the expected degrees of the nodes have to be considered, or in the general case, for any expected degree distribution, the probability of finding connected nodes at a certain angular distance presents a sharp contrast between \(D=1\) and \(D\geq2\). Then, we digress briefly on the number of neighbors on \(D\)-spheres. In the last part of the paper, we show how this affects the possibility to generate hyperbolic networks with community structure. This is quantified on block matrices representing how the generated communities are related to one another in Sec.~\ref{sec:community}.
%
%
\section{\label{sec:intro}Hyperbolic networks model}
We first review basic notions and establish some useful notation about hyperbolic random graphs in any dimension, before presenting some of their most remarkable properties. 

A hyperbolic space is a complete, simply connected, Riemannian manifold of constant negative curvature \(-\zeta^2\) and dimension \(D+1\)~\cite[Chap.~8]{Ratcliffe2006}\cite{Myers1935}. The lowest dimensional space with \(D=1\) is the hyperbolic plane, a smooth surface that can be modeled as one sheet of a hyperboloid in the three-dimensional Minkowski space, but also using other equivalent models like the upper half-plane, the Klein disk and the Poincaré disk~\cite{Ratcliffe2006, Stillwell1996}.  

Hyperbolic random graphs are based upon the hyperboloid model \(H^2\), where all points of the hyperbolic plane are parametrized using coordinates \(\varphi\in[0,\, 2\pi)\) and \(r\in[0, {R})\) and thus have a natural projection on a circle through coordinate $\varphi$~\cite{Krioukov2010}. An analogous coordinate parametrization is used in higher dimension. The angular coordinate then map points to the \(D\)-sphere instead of the circle, \emph{i.e.} \(\varphi = (\varphi_1, ..., \varphi_D)\), with \(\varphi_1,...,\varphi_{D-1}\in[0,\pi)\) and  \(\varphi_D\in[0,2\pi)\)~\cite{Yang2020}\cite[Sec. 3.4]{Ratcliffe2006}. The distance \(d_h\) between two points \(x,x'\in H^{D+1}\) whose respective coordinates are \((\varphi, r)\) and \((\varphi', r')\) is given by the hyperbolic law of cosines,
\begin{multline}\label{metric_hdp1}
  d_h(x, x') =   \frac{1}{\zeta}\mathrm{arcosh}(\cosh\zeta r\cosh\zeta r'\\ - \sinh\zeta r\sinh\zeta r'\cos\theta),
\end{multline}
where \(\zeta\in(0,\infty)\) is related to the hyperboloid's curvature \(-\zeta^2\) and \(\theta=\acos(\nicefrac{x\cdot x'}{|x|\,|x'|})\) is the angular distance between \(x\) and \(x'\), here considered as two vectors in $\mathbb{R}^{D+1}$. This generalization is referred to as the hyperboloid model \(H^{D+1}\). One important aspect of Eq.~\eqref{metric_hdp1} is that, for sufficiently large \(r,r'\), and for small \(\theta\),
\begin{equation}\label{metric_hdp1_approx}
  d_h(x, x') \approx r + r' + \frac{2}{\zeta}\log\biggl(\frac{\theta}{2}\biggr),
\end{equation}
where the first expression becomes exact in the large network limit and the last expression holds for small \(\theta\). This approximation was first published in Ref.~\cite{Krioukov2010}, but one can also refer to Refs.~\cite{Yang2020, Alanis-Lobato2016} for more detailed derivations and bounds on the approximation.\\

What is referred to as the \(\mathbb{H}^{D+1}\) model throughout this paper is not only the hyperbolic space presented above, but a random graph defined on this space. Consider a set of \(N\) nodes, \(V =\{x_i\}_{i=1,\hdots,N}\subset H^{D+1}\), where each \(x_i\) is a continuous random variable on \(H^{D+1}\). A natural choice to study the effect of hyperbolic geometry on the graph is to sample uniformly in a subset of \(H^{D+1}\), but most models can also generate networks with other node densities on the hyperbolic disk, which is reflected in the diversity of possible node degree distributions that have been studied~\cite{Serrano2008, Krioukov2010, Yang2020}. The connection probability
\begin{equation}\label{prob_edge_hyp}
  p_h(x, x') = \frac{1}{1+e^{\beta\zeta[d_h(x, x')-{\mu}]/2}}
\end{equation}
defines a random graph with node set \(V\), where each edge is an independent Bernoulli random variable with chance of success \(p_h(x, x')\). We stress that there are two levels of randomness. First, the nodes' positions are sampled from a continuous distribution on \(H^{D+1}\). Second, each realization of those positions defines a discrete probability measure on the set of all simple graphs of size \(N\). For uniformly distributed nodes, shortest paths on graphs sampled from \(\mathbb{H}^{2}\) follow the shortest paths on the underlying hyperbolic space \(H^{2}\) with high probability, a phenomenon first observed in \cite{Boguna2009b, Krioukov2010}. This phenomenon is referred to as \emph{hyperbolic routing} or \emph{congruence} between the graphs and their underlying space~\cite{Boguna2021}, and is yet to be studied in \(D>1\).

Two additional parameters are introduced in Eq.~\eqref{prob_edge_hyp}: \({\mu}>0\) that sets a connectivity distance threshold and tunes the expected average degree when nodes are sampled uniformly~\cite{Krioukov2010} and \(\beta>0\) that controls for the range of connection probabilities. Akin to the phase transition at \(\beta=1\) in the original model~\cite{Krioukov2010}, there is a phase transition at \(\nicefrac{\beta}{D}=1\), a critical value for which uniform hyperbolic random graphs have different average expected degree and power-law exponent of the degree distribution, in the asymptotic limit of large graphs~\cite{Kitsak2023, vanderKolk2021, Kovacs2022}. Our work takes place in the so-called \enquote{cold} regime, \(\beta>D\), which has been shown to generate graphs with power-law degree distributions and low average degree. It also follows that the ratio \(\nicefrac{\beta}{D}\) is kept constant when comparing models of different dimensions in Secs.~\ref{sec:dimension} and \ref{sec:community}, in order to study different dimensions in the same asymptotic regime.

With the model now explicitly defined, let us now take at closer look at which node properties the radial and angular coordinates abstract. As mentioned in the introduction, the radial coordinate encodes nodes popularity since the closer a node is to the center of the hyperbolic ball, the higher its expected degree will be~\cite{Krioukov2010, Papadopoulos2012}. The angular coordinates abstract an ensemble of similarity attributes, properties of the data underlying the network that drive nodes to be connected or not, independently of their degree. Those attributes do not have to be known explicitly, and can be related in a non-trivial way, in a manner reminiscent of how some data can be abstracted by principal components or eigenvectors. This is why detecting how many similarity dimensions would be needed to adequately model a given networked dataset is of research interest~\cite{Almagro2022,Gu2021}.

The random graph model can also be defined in the \(\mathbb{S}^D\) representation~\cite{Serrano2008}, using the same angular coordinates that maps the nodes to a \(D\)-sphere \({S^D=\{x\in\mathbb{R}^{D+1}\,|\, |x|^2=\hat{R}^2\}}\), but assigning to each node a new continuous random variable, the \emph{latent degree} \(\kappa\in(\kappa_0, \infty)\), instead of the radial coordinate \(r\). The following change of variables, a \(D\)-dimensional generalization of what was first done in \cite{Krioukov2009, Krioukov2010}, transforms from \(r\) to \(\kappa\) and inversely through 
\begin{equation}\label{ch_var}
  \kappa = \kappa_0\exp\biggl[\frac{\zeta D }{2} ({R}-r)\biggr].
\end{equation}
In \(D=1\), this simple mapping has been reinstated regularly over the last decade to highlight the correspondence between both models~\cite{Garcia-Perez2019, Boguna2021}, yet to the best of our knowledge it had not been previously published in arbitrary dimension.
Introducing one last parameter \(\hat{\mu}\), we can set \({{\mu}=(\nicefrac{2}{\zeta })\log[\nicefrac{2\hat{R}}{(\hat{\mu}\kappa_0^2)^{1/D}}]}\) in Eq.~\eqref{prob_edge_hyp} and using the hyperbolic distance approximation of Eq.~\eqref{metric_hdp1_approx}, we obtain
\begin{equation}\label{prob_edge_kappa}
  p_s(x, x') = \frac{1}{1+\left[\frac{\hat{R}\theta}{(\hat{\mu}\kappa\kappa')^{\nicefrac{1}{D}}}\right]^\beta},
\end{equation}
the connection probability as originally defined in the latent variables formalism~\cite{Serrano2008}. Albeit having different values for the equivalence between connection probabilities to hold, parameters \(\hat{R}\) and \(\hat{\mu}\) in the \(\mathbb{S}^D\) model play a similar role, respectively with regards to nodes's density on the space and the mean degree, as \(R\) and \(\mu\) in the \(\mathbb{H}^{D+1}\) model, which justifies this choice of notation~\footnote{Our notation also follows~Ref.~\cite{Boguna2021}.}.

For large sparse graphs with uniformly sampled angular coordinates, the parameter \(\hat{\mu}\) can be tuned such that the expected degree of a node with latent degree \(\kappa\) (sampled from any distribution) is proportional to the value of \(\kappa\)~\cite{Serrano2008, Garcia-Perez-phd}, where the expectation is over all possible realizations of \(\mathbb{S}^D\), hence the name \enquote{latent degree}. In addition to this relation between latent and expected degrees, Eq.~\eqref{prob_edge_kappa} highlights how angular coordinates encode similarity and latent degrees encode popularity, since the connection probability gets closer to 1 for small angular distance \(\theta\) and high latent degrees \(\kappa, \kappa'\). For this reason, the \(D\)-sphere parameterized by angular coordinates is sometimes referred to as the \emph{similarity space}. It is worth mentioning that in the literature, the notation is often simplified by adequately fixing parameters, which has not been done heretofore to keep the relationship between cited articles more accessible. 

The change of variables defined in Eq.~\eqref{ch_var} preserves the connection probabilities between nodes whenever Eq.~\eqref{metric_hdp1_approx} is valid. For uniformly distributed nodes on a ball of \(H^{D+1}\), it has been shown that the proportion of node pairs for which this is true tends to 1~\cite{Boguna2021}. Thus, random graph models within either \(\mathbb{S}^D\) or \(\mathbb{H}^{D+1}\) are considered equivalent. In our work the \(\mathbb{S}^D\) representation is used without loss of generality. 

Given latent degrees \(\kappa,\kappa'\), we can define a new continuous random variable,
\begin{equation}\label{eq:eta}
  \eta(\kappa, \kappa'):=\frac{(\hat{\mu}\kappa\kappa')^{\nicefrac{1}{D}}}{\hat{R}},
\end{equation}
whose outcome depends on the probability density function (pdf) of latent degrees, as well as parameters \(\hat{\mu}\) and \(\hat{R}\). Letting go of the explicit dependency on \(\kappa\kappa'\) for brevity, the connection probability between a pair of nodes, given by Eq.~\eqref{prob_edge_kappa}, can be written as
\begin{equation}\label{prob_edge_eta}
  p(\theta, \eta) = \frac{1}{1+(\theta/\eta)^\beta},
\end{equation}
which highlights that \(\eta\) acts as a local angular distance connectivity threshold. Indeed, 
%
\begin{equation}\label{eq:heaviside}
  \lim_{\beta\to\infty}p(\theta, \eta) = H(\eta-\theta),
\end{equation}
where \(H\) is the Heaviside step function. Thus, in the \(\beta\to \infty\) regime, all node pairs for which \(\theta<\eta\) would be connected. In the \(\mathbb{S}^D\) representation, it is common to fix the radius \(\hat{R}\) such that the number of nodes \(N\) is equal to the surface area of the \(D\)-sphere~\cite{Serrano2008, Garcia-Perez2019}, which yields a node density of 1 with
\begin{equation}\label{eq:radius}
  \hat{R} = \qty(  \frac{\Gamma(  \frac{D+1}{2}  )N}{2\pi^{ \nicefrac{(D+1)}{2} }}  )^{\nicefrac{1}{D}}.
\end{equation}
In this setting, \(\eta\) varies as \((\nicefrac{\kappa\kappa'}{N})^{1/D}\). Thus, we can think of \(\eta\) as capturing the pairwise popularity in a way reminiscent of how the angular distance \(\theta\) captures the similarity of node pairs. 

A common choice of 
pdf for angular coordinates is such that points parametrized by \(\varphi\) are uniformly distributed on \(S^D\). For latent degrees, we consider a Pareto distribution with mean~\(\bar{\kappa}\),
\begin{equation}\label{pareto_kappa_pdf}
  f_K^{\mathrm{Pareto}}(\kappa) = (\gamma-1)\,\kappa_0^{\gamma-1}\kappa^{-\gamma},\qquad\kappa_0 = \frac{\bar{\kappa}(\gamma-2)}{\gamma-1},
\end{equation}
which, with \({\gamma = 2\zeta +1}\), is akin to sampling nodes uniformly from a hyperbolic ball of \(H^{D+1}\), but offer more freedom on the shape of the degree distribution\footnote{The \(\bar{\kappa}\) is used for the mean value to avoid confusion with expected values in random graphs.}. 

\begin{figure}[t!]
\centering
\includegraphics[width=\columnwidth]{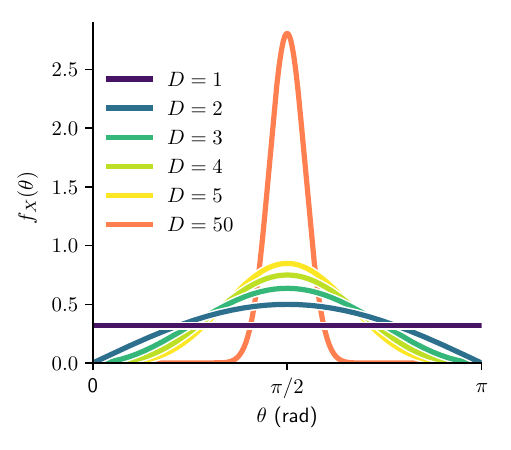}
\caption{\label{fig:distance_sphere}Probability density function of angular distance between points sampled uniformly at random on a \(D\)-sphere.}
\end{figure}

Increasing the dimension of spheres is far from being as intuitive as unfolding more dimensions of flat spaces. Let us picture this using angular distance distributions on \(D\)-spheres. Let \(X\) be a random variable describing the angular distance \(\theta\in[0,\pi]\) between two points sampled uniformly at random on \(S^D\). As derived in~\cite{Cai2013}, the pdf of \(X\) is given by 
\begin{equation}\label{distribution_angle_uniform}
  f_X(\theta) =  \frac{1}{I_D}\sin^{D-1}\theta,
\end{equation}
with 
\begin{equation}\label{I_D}I_D = \int_0^\pi\sin^{D-1}\theta\,\dd \theta = \frac{\Gamma(\nicefrac{D+1}{2})}{\sqrt{\pi}\,\Gamma(\nicefrac{D}{2})}.
\end{equation}
Figure~\ref{fig:distance_sphere} shows that for \(D\gg1\), chances are that any other point will be found at more or less an angular distance of \(\theta=\nicefrac{\pi}{2}\), a surprising property related to the concentration of measure phenomenon~\cite{Ledoux2001}. Yet even for very low \(D\), there is a significant qualitative shift between uniformity in \(D=1\) and unimodality in \(D\geq2\). This well-known property hints intuition for the upcoming section, where instead of sampling pairs of points on \(D\)-spheres, we study edges of hyperbolic random graphs.
%
%
\section{\label{sec:dimension}Effects of dimensionality}

Dimension is a fundamental property of spaces. An ant living on a circle would have much less freedom than one living on a sphere. Yet, if those two ants were to talk to each other, they might agree that they live on the same kind of space because some properties, like the possibility to come back to the same point by going straight ahead long enough, are the same. As presented in Sec.~\ref{sec:intro}, increasing the dimension of  hyperbolic random graphs boils down to considering more angular coordinates that map to the sphere or higher dimensional \(D\)-spheres, but how does this impact the graph structure? Some properties of the graphs are almost unchanged, like the degree distribution~\cite{Yang2020}, while some others, like the short cycle structure~\cite{Almagro2022}, are affected significantly. 

Our take on this question broadly deals with neighborhoods, of connected nodes in the graph but also simply of points on \(D\)-spheres. Since the dimension affects primarily the angular similarity space, the focus is first on uniform distribution of nodes on \(\mathbb{S}^D\) and the study of angular distance between nodes that are connected. On another note, we then study how the number of nearest neighbors on \(D\)-spheres varies with dimension.

\subsection{\label{sec:prob_dtheta}Angular distance between connected nodes}

In \(\mathbb{S}^1\), most edges are observed between nodes separated by a very small angular distance, except for nodes of very high expected degree (see for instance~\cite[Fig.~5.]{Krioukov2010}). We have found this propriety to change with the dimension of the underlying hyperbolic space. This is to be expected since the probability to find any other node at a very close angular distance decreases with dimension, as shown in Fig.~\ref{fig:distance_sphere}. Yet, this effect of dimensionality has interesting consequences on hyperbolic random graphs, especially when comparing \(D=1\) and \(D>1\).

Consider a \(\mathbb{S}^D\) model of \(N\) nodes as defined in Sec.~\ref{sec:intro} with angular coordinates sampled uniformly with respect to the spherical measure and latent degrees sampled from any pdf \(f_K(\kappa)\). We study the distribution of the angular distance between connected nodes within this hyperbolic random graph, as a way to assess the interplay between the dimension of the similarity space and the topology of the graph. A pairwise case of two nodes with given latent degrees is first examined, then its generalization to the whole latent degree distribution. The hard threshold limit described by Eq.~\eqref{eq:heaviside} is also studied since it allows for some insightful approximations of our results. 

\subsubsection{For a pair of nodes, general case\label{subsec:pdf_pairwise}}

\begin{figure}[b]
\centering
\includegraphics[width=\columnwidth]{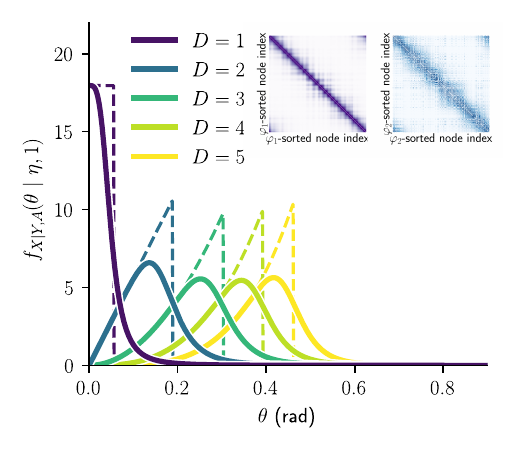}
\caption[Probability density functions of angular distance between connected nodes in the \(\mathbb{S}^D\) model.]{\label{fig:densities} Probability density functions of angular distance between connected nodes in the \(\mathbb{S}^D\) model given by Eq.~\eqref{eq:pdf}, for \(N=1000\) nodes, with \(\kappa=\kappa'=10.0\) and \(\nicefrac{\beta}{D}=3.5\). The dashed lines show the \(\beta\to \infty\) limit. Inset illustrates \(\mathbb{S}^D\) connection probability matrices for $D=1$ (left) and $D=2$ (right), \(\kappa_i=10.0\) for all nodes \(i\). The node indices were sorted to satisfy $i<j$ when $\varphi_D(i)<\varphi_D(j)$, where  $\varphi_D(i)$ denotes the coordinate $\varphi_D\in[0,2\pi)$ of the $i$-th node on the $D$-sphere.}
\end{figure}

Consider a pair of nodes with latent degrees \(\kappa, \kappa'\) such that \(\eta\) is given by Eq.~\eqref{eq:eta}. We study the pdf of angular distance between those two nodes, provided that they are connected in the random graph. Let \(X\) be the continuous random variable describing the angular distance \(\theta\in[0,\pi]\) between these nodes. Since the angular coordinates of the graph are uniformly distributed, the pdf of \(X\) is given by Eq.~\eqref{distribution_angle_uniform}. Let \(Y\) be the continuous random variable describing the value of \(\eta\), whose support depends on possible values of \(\kappa\). Finally, let \(A\) be the discrete Bernoulli random variable describing whether the nodes are connected (\(A=1\)) or not (\(A=0\)) according to the probability of Eq.~\eqref{prob_edge_kappa}. The pdf of angular distance between two nodes is given by the conditional pdf
\begin{equation}\label{def1_conditional}
f_{X|Y,A}(\theta\,|\,\eta,\, 1) =\frac{f_{X,Y,A}(\theta,\eta,1)}{f_{Y,A}(\eta,1)} 
\end{equation}
with the left-hand side notation standing more compactly for \(f_{X|Y,A}(X=\theta\,|\,Y=\eta,\, A=1)\).  Since \(A\) depends on \(\theta\) and \(\eta\) through the connection probability of Eq.~\eqref{prob_edge_kappa}, but random variables are otherwise independent, the numerator of Eq.~\eqref{def1_conditional} is given by
\begin{equation}\label{conditional_num}
f_{X,Y,A}(\theta,\eta,1) = \frac{f_X(\theta)f_Y(\eta)}{[1+(\nicefrac{\theta}{\eta})^\beta]}.
\end{equation}
Likewise, the denominator is 
\begin{equation}\label{conditional_denum}
f_{Y,A}(\eta,1) = f_Y(\eta)f_{A|Y}(1\,|\,\eta)
\end{equation}
where \(f_{A|Y}(1\,|\,\eta)\) is the following density, with \(X\) marginalized out,
\begin{multline}\label{conditional_marginal}
f_{A|Y}(1\,|\,\eta) = \int_0^\pi \frac{f_X(X)}{1+(\nicefrac{\theta}{\eta})^\beta}\dd\theta \\ = \int_0^\pi \frac{\sin^{D-1}\theta}{I_D[1+(\nicefrac{\theta}{\eta})^\beta]}\dd\theta.
\end{multline}
Altogether, this yields the explicit expression for Eq.~\eqref{def1_conditional}
\begin{equation}\label{eq:pdf}
f_{X|Y,A}(\theta\,|\,\eta,\, 1) = \frac{1}{\mathcal{Z}(\eta)}\frac{\sin^{D-1}\theta}{ [1+(\nicefrac{\theta}{\eta})^\beta] },
\end{equation}
with normalization 
\begin{equation}\label{eq:Z}
\mathcal{Z}(\eta) := \int_0^\pi \frac{\sin^{D-1}\theta}{1+(\nicefrac{\theta}{\eta})^\beta }\dd\theta.
\end{equation}
Intuitively, Eq.~\eqref{eq:pdf} is proportional to \(\sin^{D-1}\theta\) as in the distance distribution for uniformly distributed nodes on \(\mathbb{S}^D\) of Eq.~\eqref{distribution_angle_uniform}, while allowing for the influence of hyperbolic connection probability given by Eq.~\eqref{prob_edge_eta}. Its behavior for \(D \in \{1,2,3,4,5\}\) is shown in Fig.~\ref{fig:densities}, where for all dimensions the expected degree of nodes is kept the same. In Fig.~\ref{fig:densities} and Fig.~\ref{fig:densities_all_kappas}, we have set \(\nicefrac{\beta}{D}=3.5\) without loss of generality, since our results are not qualitatively affected by this parameter choice, as long as \(\nicefrac{\beta}{D}>1\).

For \(D=1\), the maximum is at \(\theta=0\), as depicted in the inset by the purple example connectivity matrix~\footnote{A \emph{connectivity} or \emph{connection probability} matrix is a convenient way to encode and illustrate pairwise connection probabilities. Each node corresponds to one column and one row, and the matrix entry is the edge probability between both nodes.}, where the dark diagonal line illustrates all of the connection probability concentrated to very small angular distances. The pdf of the angular distance separating connected nodes is strictly decreasing for \({D=1}\) and \({\beta<\infty}\), since it is proportional to the connection probability of Eq.~\eqref{prob_edge_kappa}. Hence, most edges will have nearly \({\theta=0}\) in \({D=1}\), as expected in the \(\mathbb{H}^2\) generative model~\cite{Krioukov2010} and shown by the purple curve in Fig.~\ref{fig:densities}. By contrast, for \(D>1\), this is relaxed, as exemplified by the blue curve and connectivity matrix for \(D=2\). 

Eq.~\eqref{eq:pdf} is unimodal for all \(D>1\), with a mode greater than zero and increasing with dimension. The existence and the location of the mode of Eq.~\eqref{eq:pdf} for \({D>1}\) is deduced by setting its first derivative with respect to \(\theta\) to zero to obtain
\begin{equation}\label{eq:pdf_mode_exact}
(D-1)\qty[\qty(\frac{\eta}{\theta^*})^\beta + 1 ] = \frac{\beta\tan\theta^*}{\theta^*},
\end{equation}
where \({\theta^*\in(0,\pi)}\) can be a maximum, a minimum or an inflection point. Since Eq.~\eqref{eq:pdf} is the product of a unimodal function and a decreasing function, we assume the critical point above is a maximum. It is shown in Appendix \ref{sec:pdf_appendix} that \(\theta^*=0\) iff \(D=1\). Furthermore, if \(\theta^*\ll1\) such that \(\tan\theta^*\approx\theta^*\), the solution of Eq.~\eqref{eq:pdf_mode_exact} can be closely approximated by
\begin{equation}\label{eq:pdf_mode_approx}
\theta^* \approx \eta \qty(\frac{1}{\frac{\beta}{(D-1)}  -1})^{\nicefrac{1}{\beta}}.
\end{equation}
This approximation is valid whenever the angular threshold \({\eta\sim(\kappa\kappa'/N)^{\nicefrac{1}{D}}}\) is small enough, which is the case for a significant fraction of node pairs in hyperbolic random graphs~\cite{Krioukov2010} and most observed degree distributions~\cite{Voitalov2019}. Both factors of Eq.~\eqref{eq:pdf_mode_approx} are increasing functions of the dimension \(D\). Therefore, for the vast majority of nodes having a ``reasonable'' expected degree, the most likely angular distance between connected nodes in \(\mathbb{S}^D\) is greater and greater as dimension increases. This is yet another indicator that most connected nodes will in fact be separated by an angular distance \(\theta>0\) in hyperbolic random graphs \(\mathbb{S}^D\) with \(D>1\). The expression \eqref{eq:pdf} encapsulates in a simple yet accurate way that the joint effect of nodes' angular distribution on \(D\)-spheres with the hyperbolic connection probability is qualitatively and quantitatively different across ultra-low dimensions of the model.

\subsubsection{For a pair of nodes, hard threshold limit}

An informative limiting case for the angular distance between connected nodes arises with \(\beta\to \infty\). Then, the connection probability becomes a step function given by Eq.~\eqref{eq:heaviside} with angular distance threshold \(\eta\). Equation~\eqref{eq:pdf} becomes
\begin{multline}\label{eq:pdf_heaviside}
f_{X|Y,A}(\theta\,|\, \eta, 1)\overset{\beta\to\infty}{\approx} \frac{H(\eta-\theta)\sin^{D-1}\theta}{\mathcal{Z}(\eta)}\\ \overset{\eta\ll 1}{\approx} \frac{H(\eta-\theta) \, D\theta^{D-1}}{\eta^D},
\end{multline}
where the second approximation uses the small angle property of sine. In Fig.~\ref{fig:densities}, the dashed line shows the exact value of Eq.~\eqref{eq:pdf} with \(\beta\to\infty\), but this is exactly the behavior expected from the truncated powers of \(\theta\) given in the last expression of Eq.~\eqref{eq:pdf_heaviside}. There is a sharp maximum at the threshold \(\eta\) for all \({D>1}\), which means quite counterintuitively that in this limit, most connected nodes will be  separated by their local maximal angular distance \(\eta\). This highlights a stark contrast between \(\mathbb{S}^1\) and \(\mathbb{S}^D\), \({D > 1}\), in a regime where the underlying hyperbolic geometry is the most binding to the topology of the graph because any pair of nodes that are close enough according to their degrees and model parameters shall be connected.

\subsubsection{For all latent degrees, general case}

\begin{figure}[t]
\centering
\includegraphics[width=\columnwidth]{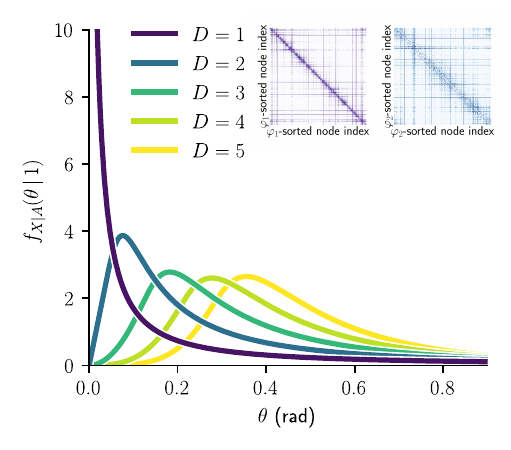}
\caption[Probability density functions of angular distance between all connected nodes in the \(\mathbb{S}^D\) model.]{\label{fig:densities_all_kappas} Probability density functions of angular distance between connected nodes in the \(\mathbb{S}^D\) model given by Eq.~\eqref{pdf_Y_cond_A}, for \(N=1000\) nodes, \(\nicefrac{\beta}{D}=3.5\) and with a Pareto latent degree distribution with \(\bar{\kappa}=10.0\) and exponent \(\gamma=2.5\). Inset show examples of \(\mathbb{S}^D\) connection probability matrices sampled with the same parameters, for $D=1$ (left) and $D=2$ (right). The node indices were sorted to satisfy $i<j$ when $\varphi_D(i)<\varphi_D(j)$, where  $\varphi_D(i)$ denotes the coordinate $\varphi_D\in[0,2\pi)$ of the $i$-th node on the $D$-sphere.}
\end{figure}

Let us now zoom out of a specific pair of nodes and study the angular distance between connected nodes considering the entire latent degree distribution. To marginalize \(\eta\) out of Eq.~\eqref{eq:pdf}, one would need to compute
\begin{equation}\label{eq:pdf_eta_init}
f_{X|A}(\theta\,|\,1) = \int_{\Omega_Y} f_{X|Y,A}(\theta\,|\,\eta,1)f_{Y|A}(\eta\,|\,1)\,\dd\eta,
\end{equation}
where the integral is computed over all possible values of \(\eta\). If we sample a graph from a \(\mathbb{S}^D\) model, and then draw an edge uniformly at random, this is precisely the pdf of the angular distance between nodes creating that edge. Using Baye's theorem, we can write
\begin{equation}\label{pdf_Y_cond_A}
f_{Y|A}(\eta\,|\,1) = \frac{f_{A|Y}(1\,|\,\eta)f_Y(\eta)}{f_A(1)}.
\end{equation}
Then, combining Eqs.~\eqref{def1_conditional}-\eqref{conditional_denum} and \eqref{pdf_Y_cond_A} yields the following explicit form for Eq.~\eqref{eq:pdf_eta_init}:
\begin{equation}\label{eq:pdf_eta}
f_{X|A}(\theta\,|\,1) = \frac{f_X(\theta)}{f_A(1)}\int_{\eta_0}^\infty\frac{f_Y(\eta)}{1+(\nicefrac{\theta}{\eta})^\beta} \,\dd\eta.
\end{equation}
To further study the angular distance between connected nodes, the marginal pdf of \(\eta\) has to be computed given the pdf of \(\kappa\), which is done in Appendix~\ref{sec:pdf_appendix}, along with the computation for \(f_A(1)\). 

Figure~\ref{fig:densities_all_kappas} illustrates the behavior of Eq.~\eqref{eq:pdf_eta} for \({D \in \{1,2,3,4,5\}}\), with expected degrees that follow the same Pareto distribution in all dimensions. The differences between \(D=1\) and \(D>1\) found in the pairwise case are still notable when we consider the whole distribution. Those are the concentration of connected nodes at very small angular distances for \(D=1\), the mode shifting towards higher angular distances with increasing dimension and the qualitative difference between the shape of distributions between \(D=1\) and \(D>1\). This effect of dimensionality, somewhat expected on its own, is to be interpreted in the light of other properties of higher-dimensional spaces studied in the following sections.

\subsection{\label{sec:neighbors}\texorpdfstring{Number of nearest neighbors}{Number of nearest neighbors}}

Dimension affects angular closeness of nodes in hyperbolic graphs, but on a more elementary level it also impacts how many points can be closest to each other on \(D\)-spheres. If a finite number \(n>2\) of points is spread on a circle, any given point will always have two, and only two, nearest neighbors. This, however, ceases to be true on higher-dimensional spheres, where the number of nearest neighbors then depends on the number of points \(n\). To quantify this, let us define a \emph{characteristic neighborhood}  \(\mathcal{B}(\phi_n)\subset S^D(\hat{R})\) as an open ball (with respect to the standard great-circle distance) on the sphere, with angular radius \(\phi_n\) of the ball chosen such that 
\begin{equation}\label{vol_Brn} 
\Vol(\mathcal{B}(\phi_n)) = \frac{\Vol(S^D(\hat{R}))}{n}.
\end{equation}
The volume refers to the \(D\)-dimensional measure of the \(D\)-sphere, i.e. the circumference of the circle, the surface area of the sphere, the volume of the \(3\)-sphere, and so on. The division of the space into areas of equal volume in Eq.~\eqref{vol_Brn} allows one to define the number of nearest neighbors \(n_\mathrm{nn}\) as
\begin{equation}\label{n_nn}
n_\mathrm{nn} = \frac{\Vol(\mathcal{B}(3\phi_n))}{\Vol(\mathcal{B}(\phi_n))} - 1.
\end{equation}
The idea is to compute the volume of an open ball of radius \({3\phi_n}\) on \(S^D(\hat{R})\) and assume that it contains \({1+n_\mathrm{nn}}\) points, a central one and its nearest neighbors. The definition of \(n_\mathrm{nn}\) is an extension of \(D=1\), where \(\mathcal{B}(\phi_n)\) is simply an arc of length \(\nicefrac{2\pi \hat{R}}{n}\). In this simplest space, we trivially have
\begin{equation}\label{n_nn_1D}
n_\mathrm{nn} = \frac{3 \phi_n \hat{R}}{\phi_n \hat{R}} -1 = 2
\end{equation}
for all \({n>2}\), in accordance with our previous intuition. 

Furthermore, open balls on \(S^2(\hat{R})\) are spherical caps of surface area \({2\pi \hat{R}^2 [1-\cos\phi_n]}\). Using Eq.~\eqref{vol_Brn} to fix the value of \(\phi_n\), it follows that for \({D=2}\),
\begin{multline}\label{n_nn_2D}
n_{\mathrm{nn}} =  \frac{n}{2} \qty[1-\cos\qty(3\acos\qty(1-\nicefrac{2}{n}))  ] -1 \\ = \frac{16}{n^2} -\frac{24}{n} +8 \ ,
\end{multline}
where we used the cosine triple angle formula.
For general dimension \(D\), \(\mathcal{B}(\phi_n)\) is a hyperspherical cap of volume 
\begin{equation}\label{vol_cap}
\Vol(\mathcal{B}(\phi_n)) = \frac{2\pi^{\nicefrac{D}{2}}\hat{R}^D}{\Gamma(\nicefrac{D}{2})}\int_0^{\phi_n} \sin^{D-1}\theta\, d \theta.
\end{equation}
It follows that
\begin{equation}\label{n_nn_D}
n_{\mathrm{nn}} = \frac{n \Gamma\qty(\frac{D+1}{2}) }{ \sqrt{\pi} \Gamma\qty(\frac{D}{2})}\int_0^{3\phi_n} \sin^{D-1}\theta\, d\theta -1,
\end{equation}
where \(\phi_n\) satisfies Eq.~\eqref{vol_Brn} using the volume of Eq.~\eqref{vol_cap}.

\begin{figure}
\includegraphics[width=0.8\columnwidth]{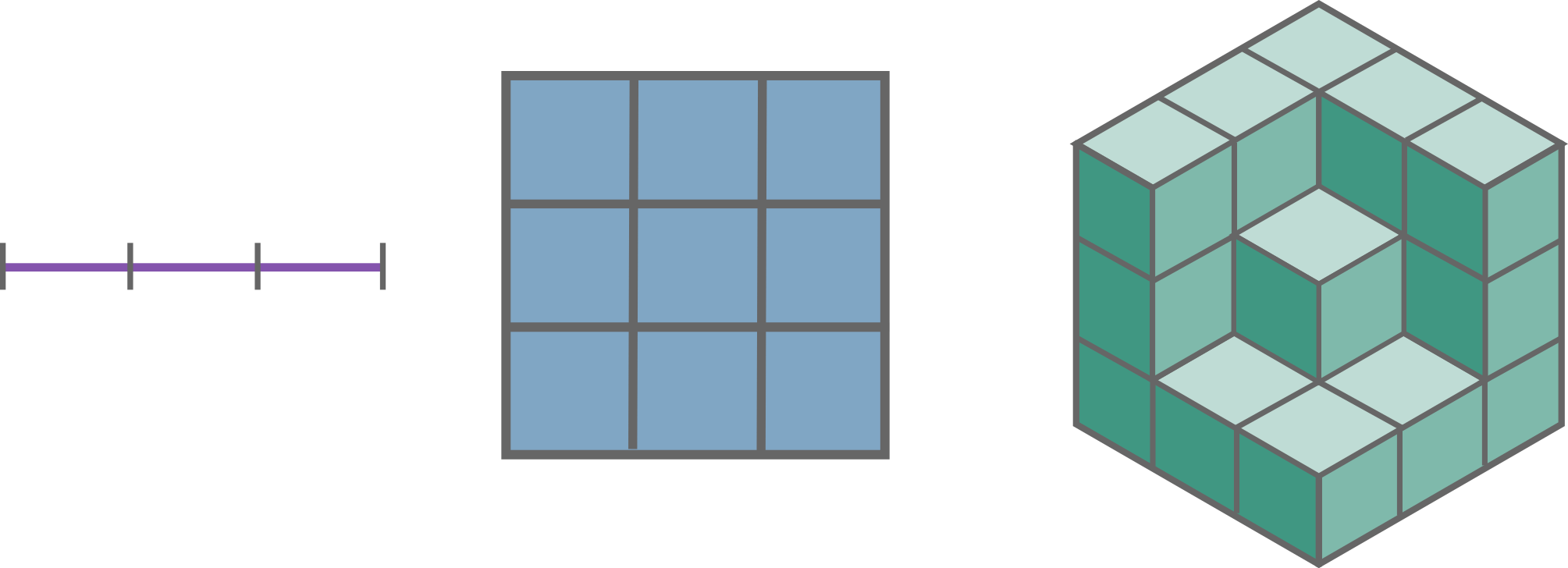}
\caption{\label{fig:pictogram}Neighboring line segments (\(D=1\)), squares (\(D=2\)) and cubes (\(D=3\)) in Euclidean space.}
\end{figure}

When \(n\gg 1\), il follows that \(\phi_n\ll 1 \). Using the approximation \(\sin x \approx x\) over the integration domains of Eqs.~\eqref{vol_cap} and \eqref{n_nn_D}, we find the asymptotic number of nearest neighbors
\begin{equation}\label{n_nn_asympt}
n_{\mathrm{nn}} \overset{n\gg 1}{\approx} 3^D-1.
\end{equation} 
In this regime, characteristic neighborhoods are equivalent to open balls in \(\mathbb{R}^D\), since \(S^D\) are manifolds, thus locally Euclidean by definition. Equation~\eqref{n_nn_asympt} can therefore be interpreted in the light of results from~\cite{Draisma2012}, where \(3^D-1\) is the maximal number of lattice directions inside a connected set of \(\mathbb{R}^D\). For \(D\in\{1,2,3\}\), the value of Eq.~\eqref{n_nn_asympt} can be pictured quite intuitively in Euclidean space as shown in Fig.~\ref{fig:pictogram}, hence justifying the apparently \emph{ad hoc} definition of \(n_{\mathrm{nn}}\). For instance, in an infinite square grid in \(\mathbb{R}^2\), one square always has \(8=3^2-1\) neighbors.

\begin{figure}[t]
\centering
\includegraphics[width=\columnwidth]{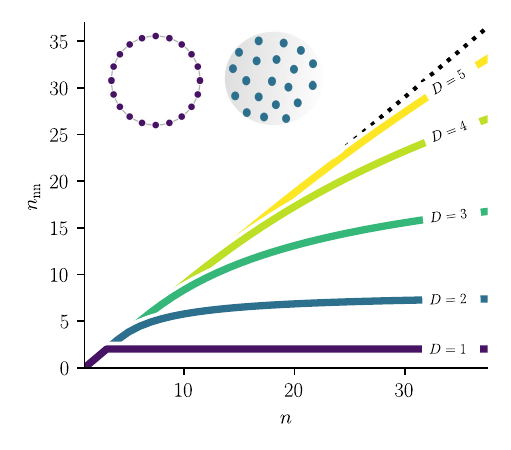}
\caption[Number of nearest neighbors \(n_{\mathrm{nn}}\) given by Eq.~\eqref{n_nn} for \(D\in\{1,2,3,4,5\}\).]{\label{fig:neighbors} Number of nearest neighbors \(n_{\mathrm{nn}}\) given by Eq.~\eqref{n_nn} for \(D\in\{1,2,3,4,5\}\). The dotted line is at \({n-1}\), the maximal number of neighbors for \(n\) points on \(S^D\). The inset shows a circle and a sphere to exemplify neighboring points in \(D=1\) and \(D=2\).}
\end{figure}

As depicted in Fig.~\ref{fig:neighbors}, \(n_{\mathrm{nn}}\) is the same in any dimension when \(n\) is low and only a few points are spread on \(S^D\). But with higher dimensions, \(n_{\mathrm{nn}}\) keeps on increasing up to higher asymptotic limits given by Eq.~\eqref{n_nn_asympt}. 

Now if, instead of counting points, \(n\) were the number of angular communities within a hyperbolic network, either real embedded networked data or a random graph model, this geometrical property of \(D\)-spheres would limit the number of communities that could be connected. The following section frame more precisely how the two effects of dimensionality we just highlighted influence community modeling within \(\mathbb{S}^D\) networks models. 
%
%
\section{\label{sec:community}Impacts on community structure}

A community is a collection of nodes that are (typically) more densely connected together than to the rest of the network, which is captured in geometric networks by a fraction of nodes that are closer together in the space. In hyperbolic networks, community structure is modeled through angular aggregation of nodes on the spherical similarity space, thus creating \emph{soft communities} \cite{Zuev2015}. In real networks embedding on \(\mathbb{S}^1\), nodes sharing qualitative attributes correlated with communities have been observed to form angular clusters. This was first observed in~\cite{Boguna2010} on the internet network, then in other types of data like economic and biological networks~\cite{Garcia-Perez2016, Garcia-Perez2019, Allard2020}. On the other hand, methods for generating modules in random hyperbolic models have also used angular closeness, either through some variant of geometric preferential attachment mechanism \cite{Zuev2015, Garcia-Perez2018} or by direct sampling of clusters as angular coordinates \cite{Muscoloni2018}. 

Dimensionality has consequences on ways in which nodes can be nearby in \(\mathbb{S}^D\), and how angular closeness is not as binding to connectivity when \(D>1\). We proceed to show how the previous findings impact community structure modeling in hyperbolic networks for ultra-low dimensions \(D\in\{1,2\}\). Numerical simulations of hyperbolic random graphs possessing community structure are at the core of the following section. Communities are generated as angular clusters and latent degrees are fixed subsequently. Once coordinates are well defined, we coarse-grain hyperbolic random graphs into block matrices that encode inter-community relations as simple weighted networks. Some global and local quantities are then measured on those block matrices to capture how communities can be related to each other in hyperbolic random graphs. This characterizes how community structure is impacted in \(\mathbb{S}^D\) at the transition between \(D=1\) and \(D=2\). 

\subsection{Generating communities in hyperbolic spaces\label{sec:sampling}}
We consider the simplest possible case where all angular communities are of similar sizes and uniformly spread on the space, with latent degrees fixed subsequently to achieve a Pareto expected degree distribution in the random graph. This allows us to experiment with simple soft community structure---yet general---random graphs that possess all relevant properties of hyperbolic networks. 
\begin{figure}[b!]
\subfloat[]{\centering\includegraphics[width=0.24\columnwidth]{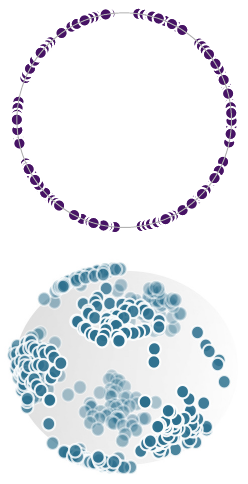}}~
\subfloat[]{\includegraphics[width=0.24\columnwidth]{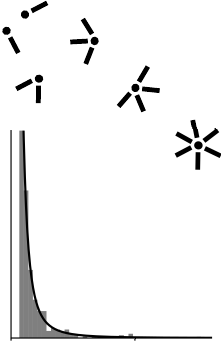}}~
\subfloat[]{\includegraphics[width=0.24\columnwidth]{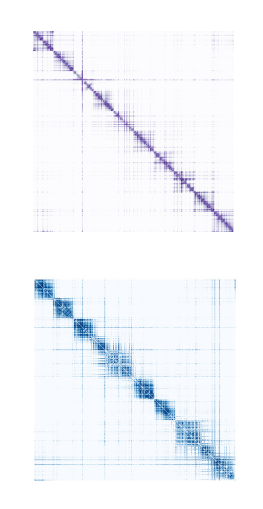}}~
\subfloat[]{\includegraphics[width=0.24\columnwidth
]{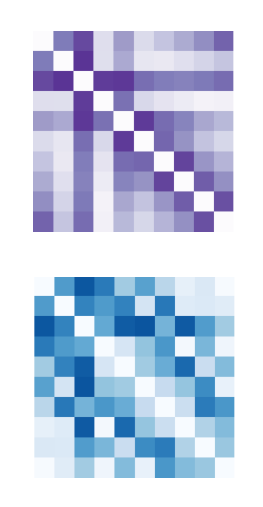}}~
\caption[Example of sampling hyperbolic random graphs with angular community structure in \(\mathbb{S}^1\) and \(\mathbb{S}^2\).]{Example of sampling hyperbolic random graphs with angular community structure in \(\mathbb{S}^1\) and \(\mathbb{S}^2\). Angular coordinates (a) are used to optimize latent degrees such that a given expected degree distribution (b) is achieved. The random graphs are then fully specified by Eq.~\eqref{prob_edge_kappa}, which is illustrated in (c) matrices of connection probabilities between all nodes, ordered with polar coordinates within each community. Those are then coarse-grained into block matrices (d) using Eq.~\eqref{eq:block_mat} to study inter-community interactions in hyperbolic spaces.\label{fig:sampling}   }
\end{figure}

Let \(n\) be the number of angular communities we wish to generate. Angular coordinates for each node are first sampled such that \(n\) clusters are distributed homogeneously on the space, as shown in Fig.~\ref{fig:sampling}(a). The disparity of nodes within each angular cluster is tuned via a parameter \(\sigma\in[0,1]\) comparable to the standard variation of normal distributions. When \(\sigma=0\), all nodes of a cluster have the same angular coordinate, whereas  when \(\sigma=1\), sampling is roughly equivalent from sampling points uniformly on \(S^D\). This procedure is similar to sampling Gaussian mixtures on the circle of Ref.~\cite{Muscoloni2018}, more details are given in Appendix~\ref{sec:sampling_appendix} and the code is freely available online~\cite{Desy2022}.

Once angular coordinates are fixed, latent degrees are optimized to obtain a Pareto expected degree distribution using the scheme of Ref.~\cite[Sec.~2.1.2]{Garcia-Perez2018}, with a tolerance of 0.2. We model independently the angular coordinates and the latent degrees, which is equivalent to assuming that the similarity space giving rise to some community structure is decoupled from the degrees of nodes within the graph. It follows from this assumption that the degree distribution within each angular cluster is drawn independently from the same distribution. With such angular coordinates and latent degrees, the hyperbolic random graph is fully defined by the connection probabilities of Eq.~\eqref{prob_edge_kappa}. Each node \({i=1,\hdots,N}\) has an additional community label \(c_i = 1,\hdots,n\) describing its membership to one of the angular clusters, which is redefined as the closest centroid. It follows that all communities are of similar size, albeit not identical, and that there is no overlap between communities, as exemplified in Fig.~\ref{fig:mosaic}(b). 

To study community structure, each random graph is mapped to a weighted graph of inter-community edges probabilities. Consider the \(n\times n\) matrix
\begin{equation}\label{eq:block_mat}
B_{uv} = \frac{1}{m} \sum_{i=1}^N\sum_{j=1}^N p_{ij}\delta(c_i, u)\delta(c_j, v)(1-\delta(c_i, c_j)),
\end{equation}
where \(m : = \sum_{i<j} p_{ij}(1-\delta(c_i, c_j))\) is the sum of probabilities associated with inter-communities edges, which can be interpreted as the total expected number of edges between distinct communities. Matrix \(B\)'s elements are normalized sums of edge probabilities between two communities, with diagonal set to zero. Thus, \(B\) can be thought of as a weighted graph describing how distinct communities interact with each other. It is normalized with the total expected number of inter-community edges such that \(B_{uv}\) quantifies the probability of finding an edge between the corresponding pair of communities \(u\) and \(v\). The complete procedure to sample a given random graph to obtain a block matrix \(B\) is illustrated in Fig.~\ref{fig:sampling}.

\subsection{Global assessment of angular dependency}

Anecdotally, blocks near the diagonal within matrices of Fig.~\ref{fig:sampling}(d) suggest that community structure in \(\mathbb{S}^1\) is impacted differently than in \(\mathbb{S}^2\). Indeed, similarly-generated hyperbolic random graphs in \(\mathbb{S}^2\) look more permissive with regards to how community blocks can be related to one another. To quantify these observations, we use the stable rank and the Shannon entropy of matrices \(B\). Both quantities are global measures of matrix structure, related in a complementary way to how diagonal versus uniform a matrix can be. 

\subsubsection{Stable rank}
The rank of a matrix is a global measure intimately related to dimensionality. In its formal definition, the rank is the maximum number of linearly independent columns or rows of a matrix, thus counting the dimension of the vector space it generates \cite{Horn2013}. When working with noisy or random matrices, it is more convenient to use the \emph{stable rank}, also called numerical rank or effective rank, and defined as
\begin{equation}\label{eq:stable_rank}
\mathrm{srank}(B) = \frac{1}{s_1^2}\sum_{i=1}^n s_i^2,
\end{equation}
where \(s_i,\,i=1,\hdots,n\) are the singular values of \(B\) in non-increasing order \cite{Rudelson2007, Vershynin2018}. The stable rank is always bounded above by the usual rank and is maximal for the identity matrix, diagonal matrices with non-zero diagonal elements, or Toeplitz matrices of the form
\[
B = \begin{pmatrix}
b&c&0 & \cdots\\
a& b& c& \ddots\\
0 & a & b & \ddots\\
\vdots & \ddots & \ddots & \ddots
\end{pmatrix},
\]
which makes it useful for quantifying to which extent matrix entries are concentrated near the diagonal. Moreover, the stable rank is invariant under any similarity transformation $B\mapsto PBP^\top$, where $P$ is a permutation matrix, thus ensuring that the value of $\mathrm{srank}(B)$ is independent of the node labelling in the graph corresponding to $B$.   

Since we compare different number of communities, hence block matrices of different order, we choose to work with the \emph{srank-to-dimension} ratio, 
\begin{equation}\label{eq:rank_to_dimension}
r(B) = \frac{\mathrm{srank}(B)}{n},
\end{equation}
a version of the stable rank normalized by its maximal possible value \cite{Thibeault2022}.
It follows that \(r(B)=1\) for a matrix of maximal rank, for instance a diagonal matrix, and \(r(B)=0\) for a null matrix. In between those extremes, \(r(B)\) captures to which extent the entries of \(B\) could be permuted to yield a diagonal matrix. In the context of community block matrices, this allows us to evaluate the complexity of connection patterns between communities. 

\begin{figure*}
\includegraphics[width=\textwidth]{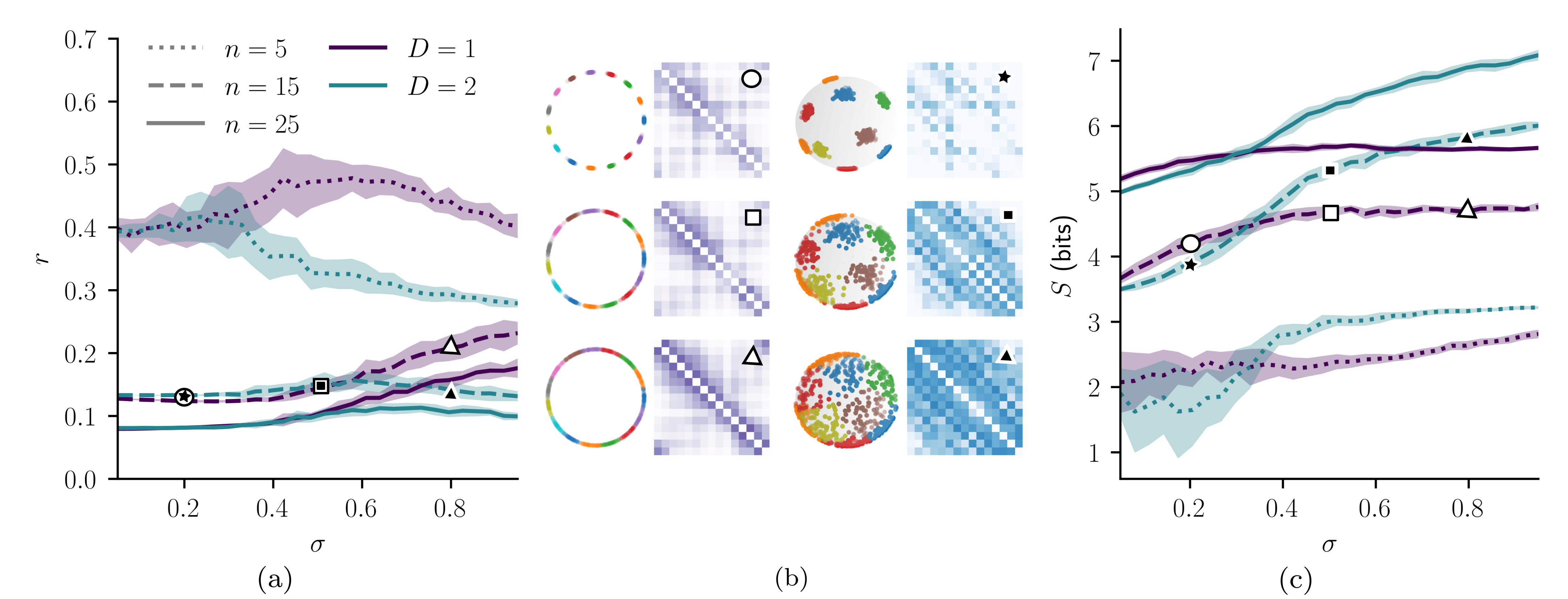}
\caption[Srank-to-dimension ratio and Shannon entropy of community block matrices in \(\mathbb{S}^1\) and \(\mathbb{S}^2\), when increasing the angular dispersion of nodes. ]{\label{fig:mosaic}(a) Stable rank to dimension ratio and (c) Shannon entropy of community block matrices in \(\mathbb{S}^1\) and \(\mathbb{S}^2\), when increasing the angular dispersion of nodes. The center panel (b) illustrates examples of angular coordinates and block matrices with \(\sigma\in\{0.2, 0.5, 0.8\}\) for \({D=1}\) (purple matrices) and \({D=2}\) (blue matrices) with corresponding values of stable rank and entropy indicated with symbols on (a) and (c). All the graphs have \(\nicefrac{\beta}{D}=3.5\) and a Pareto expected degree distribution with average degree of 4.0 and \(\gamma=2.5\).}
\end{figure*}

In Fig.~\ref{fig:mosaic}(a), we show that for various number of communities \(n\) and angular dispersion of nodes \(\sigma\), \(r\) is always higher on \(\mathbb{S}^1\) than \(\mathbb{S}^2\). Higher srank of the \(D=1\) block matrices quantifies how the inter-communities edge weights are more strictly bounded near the diagonal compared to \(D=2\). This difference is less notable when angular communities are few (\(n=5\)) and highly concentrated (\(\sigma\in[0.1, 0.2]\)), since then the additional dimension has the least impact on the neighborhood of nodes because respectively, they are mostly connected within one community. In other parameter regimes, the difference between \(\mathbb{S}^1\) and \(\mathbb{S}^2\) experimentally expresses the strong angular dependence of connectivity patterns explained in \textsection\ref{sec:prob_dtheta}, when specifically applied to community structure. 

\subsubsection{Shannon entropy}
Shannon entropy~\cite{Shannon1948} quantifies to which extent the probability mass function of a discrete random variable is uniform. It is often intuitively described as a measure of how uncertain the outcome of a random event is. As currently defined, matrices \(B\) describe the probability mass function of a single edge between two communities. It follows that the Shannon entropy of \(B\) matrices, 
\begin{equation}\label{eq:}
S(B) = - \sum_{u<v}B_{uv}\log_2(B_{uv}),
\end{equation}
quantifies how uniform block matrices are, in bits. This entropy is zero if only one entry of \(B\) has value 1, depicting a maximally non-uniform community structure. Conversely, \(S(B)\) reaches its maximal value, \(\log_2(\nicefrac{n(n-1)}{2})\), if the matrix \(B\) is entirely uniform with \(B_{uv}=\nicefrac{2}{n(n-1)}\) for all \(\nicefrac{n(n-1)}{2}\) possible community unordered pairs \((u, v)\). 

Figure~\ref{fig:mosaic}(c) shows that as nodes angular dispersion increases, block matrices in \(\mathbb{S}^2\) have an increasing entropy. This quantifies how block matrices in \(D=2\) are more and more uniform as nodes scatter across the space. Conversely, block matrices in \(\mathbb{S}^1\) uphold the same tridiagonal structure, which is reflected in the stagnation of the entropy. Again, the difference in behavior is minimal for \(n=5\), but for \(n\in\{15, 25\}\) it is not. A higher Shannon entropy of block matrices in \(D=2\), or their uniformity, means that any given community is likely to be connected to many other communities. This effect is quantified more precisely in the following section.

\subsection{Local count of neighboring communities\label{sec:community_degree}}

We also evaluate \emph{community degrees} as a hyperbolic random graph equivalent to the number of nearest neighbors \(n_\mathrm{nn}\) of Sec.~\ref{sec:neighbors}. Community degrees are computed as rows sum of a binary version of the community block matrix \(B\). We binarize matrix \(B\) through the following mapping
\begin{equation}\label{eq:binary_block_mat}
C_{uv} = \begin{cases}
1 & \text{ if } B_{uv}>\nicefrac{1}{m}, \\
0 & \hfil \text{ otherwise},
\end{cases} 
\end{equation}
where \(m\) is the sum of probabilities associated with inter-communities edges, as defined below Eq.~\eqref{eq:block_mat}. If the expected value of the number of edges between two communities is greater or equal than 1, that is to say we expect at least one edge between communities \(u\) and \(v\) on average in the random graph, then the two communities are related. This is a very liberal way to binarize \(B\); other thresholds or methods could have been used. In  Appendix~\ref{sec:binary_appendix}, we show that our numerical results for comparison of \(\mathbb{S}^1\) and \(\mathbb{S}^2\) are valid for other binarization procedures.

\begin{figure}
\centering
\includegraphics[width=\columnwidth]{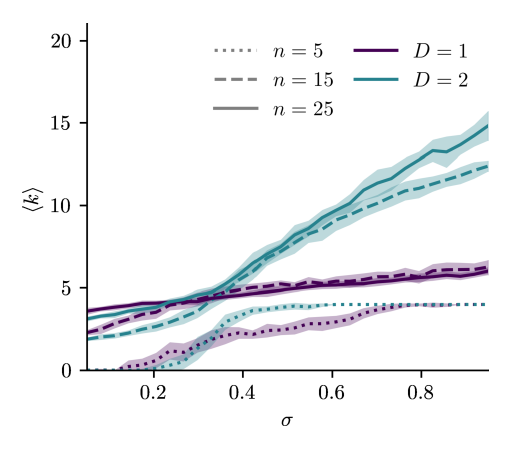}
\caption[Average community degree \(\ev{k}\) of hyperbolic random graphs with angular communities sampled in \(\mathbb{S}^1\) and \(\mathbb{S}^2\).]{\label{fig:community_degree}Average community degree \(\ev{k}\) of hyperbolic random graphs with angular communities sampled in \(\mathbb{S}^1\) and \(\mathbb{S}^2\). All the graphs have \(\nicefrac{\beta}{D}=3.5\) and a Pareto expected degree distribution with average degree of 4.0 and \(\gamma=2.5\). }
\end{figure}

The degree of community \(u\) is then defined as
\begin{equation}\label{eq:community_degree}
k_u = \sum_{v=1}^n C_{uv},
\end{equation}
which quantifies the number of other communities \(u\) is related to. We then define the \emph{average community degree} as
\begin{equation}\label{average_community_degree}
\ev{k} = \frac{1}{n}\sum_{u=1}^n k_u.
\end{equation}
In Fig.~\ref{fig:community_degree}, we show the average community degree for hyperbolic random graphs with angular communities sampled according to the scheme described in Sec.~\ref{sec:sampling}. As expected, with only \(n=5\) angular communities, the dimension has very little impact and both models reach a value of nodes dispersion \(\sigma\) where all communities are related to \(n-1\) others. Yet, when more angular clusters are considered, \(\ev{k}\) in \(\mathbb{S}^1\) barely increases,
whereas in \(\mathbb{S}^2\) communities keep on relating to more others as nodes get more dispersed on the sphere.  

An upper bound on community degree in \(D=1\) is suggested by the purple curves of Fig.~\ref{fig:community_degree}, although this is not the same phenomenon as the strict limits of Fig.~\ref{fig:neighbors}. In \textsection\ref{sec:neighbors}, the hyperbolic connection probability was not considered and the meaning of a neighborhood was purely geometric, with \(n_\mathrm{nn}\) varying with \(n\). Here, we vary the angular dispersion on nodes for a given number of communities \(n\). Yet in \(\mathbb{S}^1\), the circular boundary still poses an upper limit to how many communities can be related together. As opposed to \(n_\mathrm{nn}= 2\) in \(D=1\), the numerical upper bound of Fig.~\ref{fig:community_degree} is more than two other neighboring communities, thanks to this very permissive definition of community degree and to high degree nodes (having small radius in the hyperbolic representation) allowing for long angular range connections between communities. 

This is yet another way to assess that in \(\mathbb{S}^1\), most of the inter-community edges are concentrated to fewer nearest neighbors than in \(D=2\), as exemplified in matrices of Fig.~\ref{fig:mosaic}(b). 

\section{Summary and future work}

Recent developments in hyperbolic network geometry have challenged the common belief that only one underlying similarity dimension was enough to realistically capture the structure of real complex network~\cite{Almagro2022}. Although the impact of dimension on local properties like the degree distribution can be balanced by rescaling the abruptness of the connection probability through \(\beta\)~\cite{Yang2020, Kovacs2022, Kitsak2023}, as soon as one zoom out to mesoscopic properties like clustering or short cycles, non-trivial effects of dimension arise~\cite{Garcia-Perez2018b, Kovacs2022, Almagro2022}. Our work adds to this line of research by highlighting the interplay between dimension and community structure. We found that tighter angular bounds for connections in the \(\mathbb{S}^1\) model unrealistically restricts the community structures that can be generated on hyperbolic random graphs. Yet, dimensionality can improve current modeling of community structure in the hyperbolic framework, which has already proven successful at capturing so many other important properties all at once~\cite{Boguna2020, Boguna2021}. We found that only one additional dimension expands the inter-community connection possibilities in a way that renders more realistic modular networks modeling in hyperbolic spaces.

\begin{figure}
\includegraphics[width=\columnwidth]{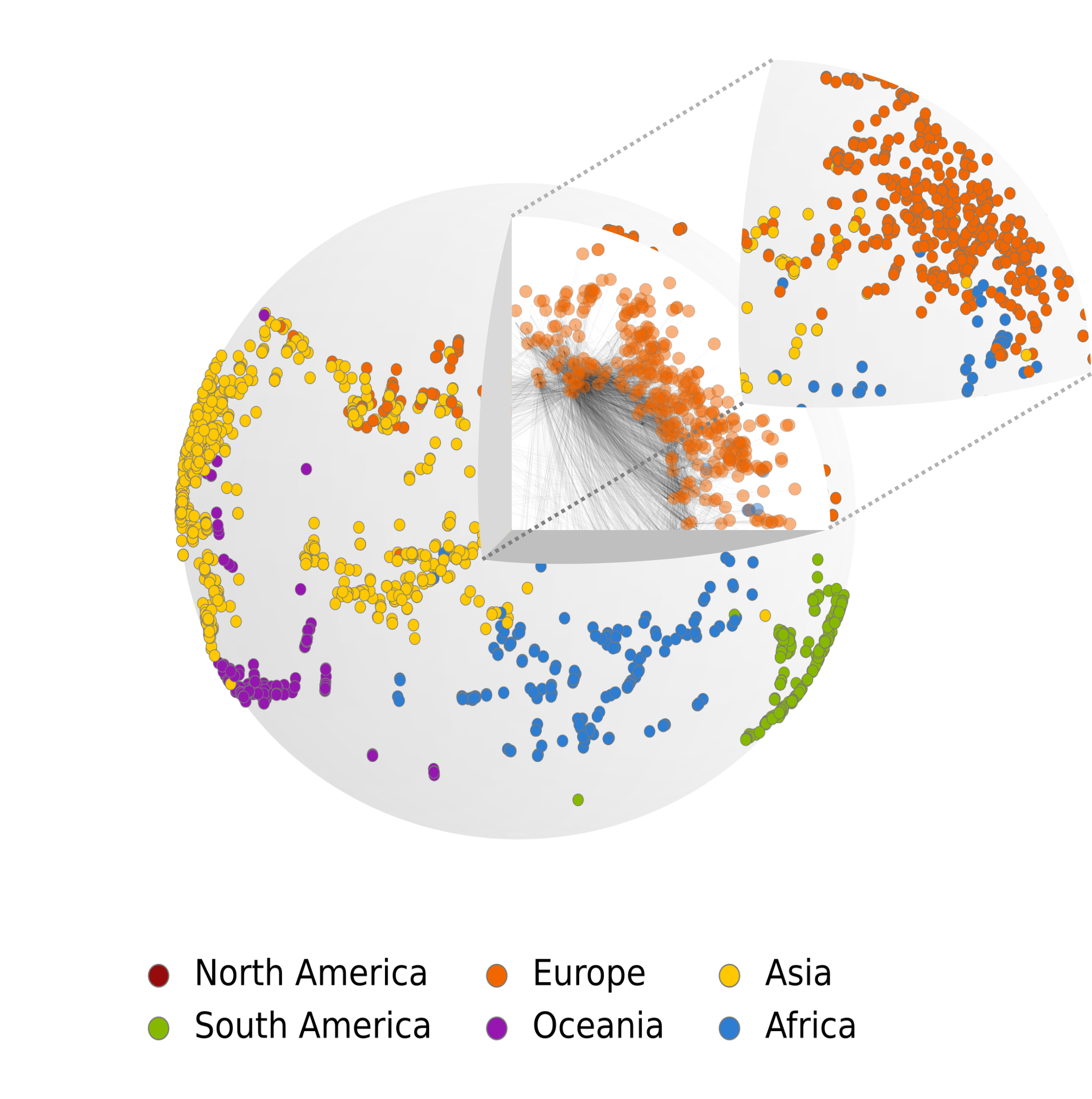}
\caption{\label{fig:S2H3} A schematic representation of the airports network from Fig.~\ref{fig:airports} embedded in \(\mathbb{S}^2/\mathbb{H}^3\). Angular coordinates of nodes projected on the surface are estimated using the algorithm from Ref.~\cite{Muscoloni2017}.  North American airports are hidden behind the sphere, see animation provided as Supplementary Material for a complete portrayal of nodes' position on \(\mathbb{S}^2\). Note that algorithm used here is based on machine learning heuristics and therefore does not maximizes the likelihood that the \(\mathbb{S}^2/\mathbb{H}^3\) model presented in Sec.~\ref{sec:intro} generated the topology of the airports network as would standard algorithms like Mercator do~\cite{Garcia-Perez2019}. The inferred angular positions are therefore used here simply to illustrate the effect dimension has on the spatial distribution of communities.}
\end{figure}

In the first part of this paper, we have shown that realized edges in hyperbolic random graphs are mostly near zero angular distance in \(D=1\), whereas it is not the case in greater dimension. This is quantified through the angular distance distribution between connected nodes given by Eq.~\eqref{eq:pdf}, which has a mode that gets further away from \(0\) as \(D\) increases. Our main result is the sharp qualitative difference between the \(D=1\) and \(D>1\) cases, which is also prominent when the angular distance distribution is averaged over all nodes' latent degrees. It follows that in \(D>1\), connection patterns between individual nodes are less restricted by their angular proximity. Besides, the number of nearest neighbors for points on spherical manifolds is also an increasing function of the dimension. Thus, the number of nearest angular neighboring clusters of nodes, or soft communities in hyperbolic random graphs, varies with \(D\), which reflects on community structure modeling. 

To assess these effects, in the second part of this paper, we experimented numerically with hyperbolic random graphs in which nodes' angular coordinates were grouped in soft communities in \(D=1\) and \(D=2\). By averaging the individual connection probabilities into the probabilities to find an edge between two distinct communities, we obtained block matrices describing inter-community relationships. The structure of these were studied across different values of angular dispersion of nodes to highlight differences arising from adding only one similarity dimension. Indeed, block matrices were more bounded to a diagonal shape by angular closeness in \(D=1\), as quantified by their stable rank, and more uniform in \(D=2\), as quantified by their Shannon entropy. The idea of having more uniform inter-community block matrices refers to the fact that in \(D>1\), any given community can be related to more other communities, as also quantified by the average community degree. Akin to the number of nearest neighbors for points, the average community degree was higher in \(D=2\), especially as soft communities were more dispersed angularly. 

In \(D=1\), mechanisms underlying soft-community formation and modeling in \(\mathbb{S}^1\) and \(\mathbb{H}^2\)~\cite{Zuev2015, Garcia-Perez2018, Muscoloni2019} and the inherent modularity of hyperbolic random graphs~\cite{Kovacs2021, Chellig2022} have been studied. However, we have argued here that hyperbolic network models of varying dimensions are quite distinct when it comes to their potential to model community structure, in particular between \(D=1\) and \(D>1\). 
Community structures as simple as a triplet of strongly interacting communities are poorly captured in \(D=1\), as exemplified by the African, Asian, and European airports in Fig.~\ref{fig:airports}, whereas the same overlapping communities become disentangled in \(D=2\), as shown in Fig.~\ref{fig:S2H3} and in the animation provided as Supplementary Material. The advantage of $D>1$ over $D=1$ for representing  community network data was also reported in Ref.~\cite{Muscoloni2019}.
Since most real-world networks possess some sort of nodes' aggregates, this suggests that the recently discovered higher underlying hyperbolic dimension for most real-world networks by Ref.~\cite{Almagro2022} could be related to their mesoscale structure, although dimension detection is beyond the scope of our paper and should be explored in future work. 

Increasing the dimensionality of the underlying hyperbolic spaces might also help to improve the likelihood maximization procedure that is used for inferring hyperbolic coordinates for networked data. As observed for three different embedding algorithms~\cite{Papadopoulos2015, Muscoloni2017, Garcia-Perez2018}, inferring angular coordinates can be considered the hardest part of the hyperbolic embedding procedure. The use of common neighbors has been considered in Ref.~\cite{Papadopoulos2015b} to solve this issue. We have shown that the number of nearest angular neighbors increases with dimension, as does the diversity of community structures that can be generated. Hence our conjecture that higher, albeit still ultra-low, dimensional hyperbolic spaces would more realistically capture the structure of networks and be inferred without such angular degeneracy. 

Another way to see this is with a simple thought experiment. Let us assume there exist some networks whose topology is naturally reflected by a higher underlying hyperbolic dimension. For instance, one could think of a randomly generated network in \(\mathbb{H}^{D+1}\) with \(D>1\), or a real network like the internet which, according to Ref.~\cite[Fig.~5]{Almagro2022} could have a dimension as high as \(D=7\). If such a network were to be embedded in \(\mathbb{H}^2\), the geometric neighborhoods of nodes (which node is close to which  others) could not be respected, since as we show in Sec.~III.B. the number of nearest angular neighbors varies with dimension. Therefore, any lower-dimensional embedding will have to accommodate by positioning some nodes closer than they should and vice versa, which would reflect on tasks like link prediction, for instance. We wish we could carry out this experiment, but current embedding algorithms are yet to be generalized to more dimensions~\cite{Garcia-Perez2019}, or have only been used on relatively small networks~\cite{Muscoloni2017, Muscoloni2019} in \(D=2\).

Knowledge about community structure has already been shown to impact performance of very diverse network tasks, from hyperbolic embedding coordinate inference~\cite{Faqeeh2018, Wang2016a,Wang2016b} to dimension reduction~\cite{Jiang2018, Thibeault2020,Vegue2022}, efficient information communication~\cite{Lynn2020} and resilience~\cite{Dong2018}. A natural way to push our work will then be to study how this coupling between community structure and the underlying dimension of hyperbolic networks affects tasks which have already been shown to perform better in the framework of hyperbolic network geometry. In particular, greedy routing of information propagation has been one of the first and most interesting assets of hyperbolic geometry~\cite{Boguna2009b, Allard2020}, which should be studied in higher dimensions while explicitly considering community structure. 

Another intriguing direction of research would be to use hierarchical network generation mechanisms like geometric branching growth~\cite{Zheng2021} in higher dimensions. Such a procedure generates a hyperbolic network through subdivision of a small initial network into finer and finer descendants, a statistical inverse of geometric renormalization~\cite{Garcia-Perez2018b}. By placing the initial seed network into higher-dimensional spaces, one could compare hierarchical community structure in different dimensions. This would be akin to our numerical study but in a more complex setting where the nodes angular coordinates follow a hierarchical structure.

\begin{acknowledgments}
This work was supported by the Fonds de recherche du Qu\'ebec -- Nature et technologies, the Natural Sciences and Engineering Research Council of Canada, and the Sentinelle Nord program of Universit\'e Laval, funded by the Canada First Research Excellence Fund. We acknowledge Calcul Québec and Compute Canada for their technical support and computing infrastructures.
\end{acknowledgments}

\section*{Code availability}
The code used to produce all figures is written in the Python programming language and is available on Zenodo and Github~\cite{Desy2022}.

\section*{Author contribution}
B.D., P.D and A.A designed the research, as well as analyzed and interpreted the results. B.D. and A.A. wrote the codes for the numerical experiments. B.D. performed the research and wrote the manuscript. All authors read, commented, edited, and approved the final version of the manuscript.

\newpage


%

\newpage

\appendix

\section{Explicit computations for angular densities\label{sec:pdf_appendix}}

\subsection{Modes of angular distance between connected node pairs}

Taking the first derivative of Eq.~\eqref{eq:pdf} with respect to \(\theta\) and setting it to zero yields
\begin{equation}\label{app:1}
\frac{(D-1)\sin^{D-2}\theta^*\cos\theta^*}{1+(\nicefrac{\theta^*}{\eta})^\beta} = \frac{\sin^{D-1}\theta^*}{\qty[1+(\nicefrac{\theta^*}{\eta})^\beta]^2}\frac{\beta(\nicefrac{\theta^*}{\eta})^\beta}{\theta^*}, 
\end{equation}
where \(\theta^*\) can be a minimum, a maximum or an inflexion point. We separate two cases:
\begin{enumerate}
\item If \(\theta^*\in(0,\pi)\), Eq.~\eqref{app:1} can be simplified to Eq.~\eqref{eq:pdf_mode_exact}.
\item We proceed to show that \(\theta^*=0\) iff \(D=1\) in the regime \(\beta>D,\, \eta>0\). First, if \(D=1\), Eq.~\eqref{app:1} reduces to
\begin{equation}\label{app:1_1D}
0 = \frac{\beta(\nicefrac{\theta^*}{\eta})^\beta}{\theta^*\qty[1+(\nicefrac{\theta^*}{\eta})^\beta]^2},
\end{equation}
which is verified only for \(\theta^*=0\). Conversely, setting \(\theta^*=0\) in Eq.~\eqref{app:1} trivially yields \(D=1\).
\end{enumerate}

\subsection{\texorpdfstring{Probability density function for \(\eta\)}{Probability density function for eta}}

Since \(\eta\) is a scalar function of \(\kappa,\kappa'\), the pdf for \(Y\) can be computed as follows. 
\begin{equation}\label{pdf_eta_1}
f_Y(\eta ) = \int_{\Omega_K}\int_{\Omega_K} \rho_K(\kappa)\rho_K(\kappa')\delta(g(\kappa'))\,\dd\kappa\,\dd\kappa'
\end{equation}
with \(g(\kappa') := \eta-\nicefrac{(\mu\kappa\kappa')^{\nicefrac{1}{D}}}{R}\) and \(\delta\) the Dirac delta function. By definition of composition with the Dirac delta function under the integral, 
\begin{equation}\label{pdf_eta_2}
\delta(g(\kappa')) = \frac{\delta(\kappa'-\kappa'_*)}{|g'(\kappa'_*)|},
\end{equation}
with \({\kappa'_* = \nicefrac{(\eta R)^D}{\mu\kappa}}\), the unique root of \(g\). Hence
\begin{equation}\label{annexe0}
\dv{g}{\kappa'}\eval_{\kappa'=\kappa'_*} = \frac{\mu\kappa\eta^{(1-D)}}{DR^D}>0, 
\end{equation}
and it follows that
\begin{equation}\label{annexe1}
\delta(g(\kappa')) = \frac{\delta(\kappa'- \nicefrac{(\eta R)^D}{\mu\kappa})}{\mu\kappa\eta^{(1-D)}}DR^D.
\end{equation}
Eq.~\eqref{pdf_eta_1} can thus be computed as
\begin{equation}\label{pdf_eta_3}
f_Y(\eta ) = \frac{D R^D}{\mu\eta^{1-D}} \int_{0}^\infty \rho_\kappa(\kappa) \rho_\kappa(\nicefrac{(\eta R)^D}{\mu\kappa})\frac{ d\kappa}{\kappa}.
\end{equation}
For any distribution of \(\kappa\) with a non-zero lower bound  \(\kappa_0\),
\begin{equation}\label{regime_integrand_null}
\frac{(\eta R)^D}{\mu\kappa}<\kappa_0\quad \iff\quad \frac{(\eta R)^D}{\mu\kappa_0}<\kappa,
\end{equation}
which means that the integrand in Eq.~\eqref{pdf_eta_3} is null in the regime of Eq.~\eqref{regime_integrand_null}. We finally have
\begin{equation}\label{pdf_eta}
f_Y(\eta) = \frac{D R^D}{\mu\eta^{1-D}} \int_{\kappa_0}^{\nicefrac{(\eta R)^D}{\mu\kappa_0}} \rho_\kappa(\kappa) \rho_\kappa(\nicefrac{(\eta R)^D}{\mu\kappa})\frac{ d\kappa}{\kappa}.
\end{equation}
This pdf can be computed exactly for \(\kappa\) drawn from a Pareto distribution with parameter \(\gamma\). Let 
\begin{equation}\label{pdf_kappa_pareto}
\rho_\kappa(\kappa) = (\gamma-1)\kappa_0^{\gamma-1}\kappa^{-\gamma},
\end{equation}
then computation of the integral in Eq.~\eqref{pdf_eta} gives
\begin{equation}\label{pdf_eta_pareto}
f_Y^{\mathrm{Pareto}}(\eta) = \frac{D (\gamma-1)^2 \kappa_0^{2(\gamma-1)}\mu^{\gamma-1}} {R^{D(\gamma-1)}\eta^{D(\gamma-1)+1}}\log\qty[\frac{(\eta R)^D}{\mu\kappa_0^2}]. 
\end{equation}
\subsection{Marginalized connection probability}
Since the connection probability of hyperbolic random graphs (Eq.~\eqref{prob_edge_kappa}) depends on \(\theta\) and \(\eta\), one would need to know the pdf for \(\theta\) and the pdf for \(\eta\) to compute the marginalized connection probability that appears in the normalization of Eq.~\eqref{eq:pdf_eta}. 
\begin{equation}\label{marg_prob_edge_init}
f_A(1) = \int_{\Omega_Y}\int_{\Omega_X} f_{X,Y,A}(\theta,\eta,1)\,\dd\theta\,\dd\eta.
\end{equation}
Given \(\Omega_X=[0,\pi]\), \(\Omega_Y=[\eta_0, \infty)\) and Eq.~\eqref{conditional_num},
\begin{align}
f_A(1) & = \int_{\eta_0}^\infty \int_0^\pi \frac{\sin^{D-1}\theta\, f_Y(\eta)}{I_D [1+(\nicefrac{\theta}{\eta})^\beta]}\,\dd\theta\,\dd\eta\\
& = \frac{1}{I_D}\int_{\eta_0}^\infty\qty[\int_0^\pi \frac{\sin^{D-1}\theta}{1+(\nicefrac{\theta}{\eta})^\beta}\,\dd\theta]f_Y(\eta)\,\dd\eta\\
& = \frac{\sqrt{\pi}\Gamma(\nicefrac{D}{2})}{\Gamma(\nicefrac{D+1}{2})} \int_{\eta_0}^\infty \mathcal{Z}(\eta)f_Y(\eta)\,\dd\eta,
\end{align}
where the angular distance distribution of Eq.~\eqref{distribution_angle_uniform} and the definition of Eq.~\eqref{eq:Z} have been used.

\section{Sampling methods\label{sec:sampling_appendix}}

To sample clusters of angular coordinates for hyperbolic random graphs in \(D=2\), we first distribute the modes of all \(n\) clusters evenly using the Fibonacci lattice algorithm~\cite{Gonzalez2009}. Then, coordinates of nodes within each cluster are sampled from a three-dimensional normal distribution in \(\mathbb{R}^3\) centered around its mode, and then projected on the unit sphere. Lastly, the community to which each node belongs is defined as the one of the closest mode, or as the closest centroid. This allows us to obtain a given number \(n\) of evenly distributed clusters of angular coordinates for the nodes.

\section{Other thresholding methods for block matrices\label{sec:binary_appendix}}
\begin{figure}
\centering
\includegraphics[width=\columnwidth]{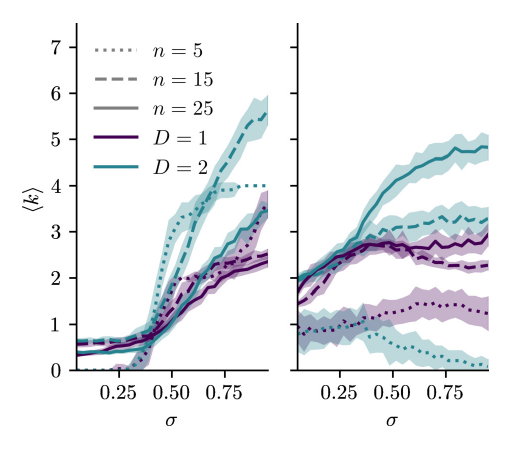}
\caption[Community degree \(\ev{k}\) with alternative binarization methods.]{\label{fig:binarizations} Community degree \(\ev{k}\) with alternative binarization methods.  On the left, Eq.\eqref{eq:community_degree} is used with a higher threshold of \(10/m\) and on the right, the disparity filter of Ref.~\cite{Serrano2009} is applied with \(\alpha=0.2\).}
\end{figure}

Here we validate that results about community degree of Sec.~\ref{sec:community_degree} are robust to other binarization methods. In Fig.~\ref{fig:binarizations}, we show the community degree \(\ev{k}\) measured on the same matrices as the ones used for Fig.~\ref{fig:community_degree}, but using a higher threshold (left) and the disparity filter of Ref.~\cite{Serrano2009} (right) to transform the inter-community edges probability matrix \(B\) to a binary matrix. Both plots show that community degree in \(\mathbb{S}^2\) increases to higher values than in \(\mathbb{S}^1\). The disparity filter penalizes locally homogeneous edge weights, which explains why the blue curves on the right panel do not increase as much as in Fig.~\ref{fig:community_degree}, since then the inter-community edge probability becomes more and more homogeneous on the neighboring angular clusters.

\end{document}